\documentclass[twoside,11pt]{article}

\usepackage[accepted]{melba}
\usepackage{mwe}

\usepackage{amsmath,amsfonts}

\usepackage{multirow}
\usepackage{makecell}
\usepackage[hypcap=false]{caption}
\usepackage{placeins}
\usepackage{amsmath,amsfonts,amssymb}
\usepackage{graphicx} 
\usepackage{multirow}
\usepackage{rotating}
\usepackage{url}
\usepackage{subcaption}
\usepackage{makecell}
\usepackage{placeins}

\melbaid{2024:012}  
\doi{https://doi.org/10.59275/j.melba.2024-7fgd}
\melbaauthors{}  
\volume{2}
\firstpageno{888}  
\melbayear{2024}  
\datesubmitted{11/2023}  
\datepublished{06/2024}  

\ShortHeadings{Automated MRI Quality Assessment in Clinical Data Warehouses:}{Loizillon et al. }

\title{Automated MRI Quality Assessment of Brain T1-weighted MRI in Clinical Data Warehouses: A Transfer Learning Approach Relying on Artefact Simulation}

\author{\firstname Sophie Loizillon \\  
	\addr Sorbonne Universit\'{e}, Institut du Cerveau – Paris Brain Institute - ICM, CNRS, Inria, Inserm, AP-HP, H\^{o}pital de la Piti\'{e}-Salp\^{e}tri\`{e}re, Paris 75013, France
	\AND
	\name Simona Bottani\\
	\addr Sorbonne Universit\'{e}, Institut du Cerveau – Paris Brain Institute - ICM, CNRS, Inria, Inserm, AP-HP, H\^{o}pital de la Piti\'{e}-Salp\^{e}tri\`{e}re, Paris 75013, France
 	\AND
	\name St\'{e}phane Mabille\\
	\addr Sorbonne Universit\'{e}, Institut du Cerveau – Paris Brain Institute - ICM, CNRS, Inria, Inserm, AP-HP, H\^{o}pital de la Piti\'{e}-Salp\^{e}tri\`{e}re, Paris 75013, France
  	\AND
	\name Yannick Jacob\\
	\addr AP-HP, WIND department,Paris 75012, France
   	\AND
	\name Aur\'{e}lien Maire\\
	\addr AP-HP, WIND department,Paris 75012, France
        \AND
	\name Sebastian Str\"{o}er  \\
	\addr AP-HP, Hôpital de la Piti\'{e} Salp\^{e}tri\`{e}re, Department of Neuroradiology, Paris 75013, France
        \AND
	\name Didier Dormont  \\
	\addr Sorbonne Universit\'{e}, Institut du Cerveau – Paris Brain Institute - ICM, CNRS, Inria, Inserm, AP-HP, H\^{o}pital de la Piti\'{e}-Salp\^{e}tri\`{e}re, DMU DIAMENT, Paris 75013, France
        \AND
	\name Olivier Colliot  \\
	\addr Sorbonne Universit\'{e}, Institut du Cerveau – Paris Brain Institute - ICM, CNRS, Inria, Inserm, AP-HP, H\^{o}pital de la Piti\'{e}-Salp\^{e}tri\`{e}re, Paris 75013, France
         \AND
	\name Ninon Burgos \email ninon.burgos@cnrs.fr  \\
	\addr Sorbonne Universit\'{e}, Institut du Cerveau – Paris Brain Institute - ICM, CNRS, Inria, Inserm, AP-HP, H\^{o}pital de la Piti\'{e}-Salp\^{e}tri\`{e}re, Paris 75013, France
          \AND
	\name for the Alzheimer's Disease Neuroimaging Initiative  \\
        \addr Data used in preparation of this article were obtained from the Alzheimer's Disease Neuroimaging Initiative (ADNI) database (\url{adni.loni.usc.edu}). As such, the investigators within the ADNI contributed to the design and implementation of ADNI and/or provided data but did not participate in analysis or writing of this report. A complete listing of ADNI investigators can be found at: \url{http://adni.loni.usc.edu/wp-content/uploads/how_to_apply/ADNI_Acknowledgement_List.pdf}.
          \AND
	\name the APPRIMAGE Study Group  \\
        \addr Members of the APPRIMAGE study group can be found at \url{https://www.aramislab.fr/apprimage}
}

\begin{document}

\maketitle

\begin{abstract}
	The emergence of clinical data warehouses (CDWs), which contain the medical data of millions of patients, has paved the way for vast data sharing for research. The quality of MRIs gathered in CDWs differs greatly from what is observed in research settings and reflects a certain clinical reality. Consequently, a significant proportion of these images turns out to be unusable due to their poor quality. Given the massive volume of MRIs contained in CDWs, the manual rating of image quality is impossible. Thus, it is necessary to develop an automated solution capable of effectively identifying corrupted images in CDWs. 
 
    This study presents an innovative transfer learning method for automated quality control of 3D gradient echo T1-weighted brain MRIs within a CDW, leveraging artefact simulation. We first intentionally corrupt images from research datasets by inducing poorer contrast, adding noise and introducing motion artefacts. Subsequently, three artefact-specific models are pre-trained using these corrupted images to detect distinct types of artefacts. Finally, the models are generalised to routine clinical data through a transfer learning technique, utilising 3660 manually annotated images. The overall image quality is inferred from the results of the three models, each designed to detect a specific type of artefact. Our method was validated on an independent test set of 385 3D gradient echo T1-weighted MRIs.
    
    Our proposed approach achieved excellent results for the detection of bad quality MRIs, with a balanced accuracy of over 87\%, surpassing our previous approach by 3.5 percent points.  Additionally, we achieved a satisfactory balanced accuracy of 79\% for the detection of moderate quality MRIs, outperforming our previous performance by 5 percent points. Our framework provides a valuable tool for exploiting the potential of MRIs in CDWs.
\end{abstract}


\section{Introduction}
Quality control (QC) is the first crucial step when working with routine clinical images from a clinical data warehouse (CDW). CDWs contain the medical data of millions of patients, such as medical reports, biological results and imaging studies. The Assistance Publique-Hôpitaux de Paris (AP-HP), the university hospital trust of the Greater Paris area, has developed its own CDW gathering more than 25 million images, including more than 200,000 gradient echo 3D brain T1-weighted (T1w) magnetic resonance images (MRIs), acquired in 39 different hospitals. This anatomical sequence is best suited for evaluating regional volume loss and is for example part of the protocol for assessing neurodegenerative diseases. MRIs are susceptible to a wide range of artefacts, resulting from both the intrinsic characteristics of the scanner and the patient's interaction with it~\citep{krupa2015artifacts}. In a previous study performed on the AP-HP CDW ~\citep{bottaniAutomaticQualityControl2021}, we showed that more than 30\% of a large representative subset of 3D T1w brain MR images were of poor quality and 25\% were completely unusable and unsuitable for further analysis. These images were affected by various common defects such as poor contrast (41\% of the images), motion (28\% of the images) and noise (22\% of the images). Given the enormous volume of images within the CDW, it is imperative to implement an effective automated QC process.

Unfortunately, due to the large number of poor quality MRIs and the presence of contrast-enhanced images (gadolinium injected T1w MRI), it is not possible to use QC solutions that require significant pre-processing, such as MRIQC~\citep{estebanMRIQCAdvancingAutomatic2017} or the approach used for the UKBioBank~\citep{ALFAROALMAGRO2018400}. These solutions rely on metrics calculated from segmentation maps, which cannot be generated accurately for many images acquired in clinical routine. Machine learning-based QC methods seem more appropriate for our problem as they do not require extensive pre-processing. For example, ~\citep{kustner2018machine} proposed a new machine learning-based reference-free MRI quality assessment framework that uses active leaning to efficiently estimate quality classes on a 5-point Likert scale. More recently, ~\cite{sujit2019automated} used convolutional neural networks (CNNs) for automated QC of T1w MRIs thanks to their ability to learn features without prior knowledge of which ones are the most suitable. However, most of these works limits their application to research images acquired following a well-defined research protocol acquisition. Other works focus on just detecting a single type of artefact, such as noise \citep{MANJON201535,kidoh2020deep} or motion~\citep{oksuzBrainMRIArtefact2021, FANTINI2021101897, VAKLI2023102850}. In a previous study, we introduced a framework for performing automatic QC of T1w brain MRIs within a CDW using fully supervised deep learning techniques~\citep{bottaniAutomaticQualityControl2021}. 
While our approach demonstrated good accuracy in identifying bad quality images, its performance in detecting medium quality images was less satisfactory, with a balanced accuracy of 74\%. In addition, we encountered difficulties in effectively detecting certain features, such as motion artefacts and moderate contrast.

A way to improve our previous QC framework would be to manually annotate more images, but this is very time consuming. An alternative consists in using simulated data coupled with a transfer learning strategy. By intentionally introducing synthetic artefacts into MR images, we have the opportunity to build a dataset as large as necessary with reliable quantitative ground truths that could be used for pre-training purposes. 
Numerous efforts have been made in the synthesis of motion~\citep{mohebbianClassifyingMRIMotion2021a, pawar2022suppressing, shaw2021decoupled, Loizillon2023TransferLearning,sagawaEvaluationMotionArtifacts2022} and noise~\citep{aja2016statistical,collins1998design} artefacts, while the literature on contrast generation is relatively limited and mainly used for data augmentation purposes~\citep{chlapReviewMedicalImage2021, perez-garciaTorchIOPythonLibrary2021}.

In this paper, we propose a novel transfer learning framework for the automatic quality control of 3D gradient echo T1w brain MRIs within a CDW based on artefact simulation. To achieve this, we corrupted images from research-oriented datasets by making the contrast poorer, adding noise and introducing motion artefacts. This process allows us to pre-train three artefact-specific models capable of detecting these different types of artefacts. We then generalised our models to 3D gradient echo T1w clinical routine MRIs using transfer learning, relying on the manual annotation of 3660 images. Finally, the overall image quality is determined based on the outcomes of the three models, each tailored to detect a specific type of artefact.

Related works on motion artefact detection was accepted for publication in the proceedings of the SPIE Medical Imaging 2023 conference \citep{Loizillon2023TransferLearning} and at  MedIA~\cite{loizillon2024automatic}. The main contribution of this paper is the extension of the transfer learning approach, initially restricted to the detection of motion artefacts, to other types of artefacts (poor contrast and noise). This allows rating the overall image quality. Additionally, we investigated the use of reliable metrics to quantify the different artefacts and validated our approach on two different tasks: the detection of poor quality and moderate quality images. Finally, we carried out a comparative analysis of two approaches: direct and indirect classification of overall image quality.

\section{Materials}
We used two different types of dataset: a research-oriented dataset and a clinical routine dataset. Initially, we worked with three publicly accessible research datasets to pre-train CNNs using images corrupted with synthetic artefacts. Subsequently, we leveraged clinical routine images from the AP-HP CDW for the purposes of transfer learning and validation.

\subsection{Research-oriented dataset}
We utilised the ADNI, MSSEG and MNI BITE databases, together denoted as our research-oriented dataset, for synthesising artefacts and pre-training our models. We emphasised the presence of contrast-enhanced T1w MRIs, as the CDW includes images obtained both with and without contrast agent. 

The Alzheimer's Disease Neuroimaging Initiative (ADNI) is a multi-site study of elderly individuals with normal cognition, mild cognitive impairment, or Alzheimer's disease~\citep{petersenAlzheimerDiseaseNeuroimaging2010}. 
In the ADNI-1 phase, more than 5000 T1w MRI scans were obtained using 1.5 T machines provided by various manufacturers, including GE, Siemens, and Philips. Out of the available scans, we selected the 1143 MRIs from 70 subjects that successfully cleared both levels of quality control ensuring compliance with protocol parameters and series-specific quality criteria, and that exhibited no motion artefacts based on the IPMOTION score.

The MSSEG MICCAI challenge, designed to achieve the segmentation of multiple sclerosis lesions, encompasses a dataset of 53 adult patients spanning across four distinct sites, as detailed in~\citep{commowick2018objective}. Four MRI scanners were employed for data acquisition: GE Discovery 3T, Philips Ingenia 3T, Siemens Aera 1.5T, and Siemens Verio 3T. Each scan within this challenge includes four distinct MRI sequences: 3D FLAIR, 3D T1w, 3D contrast-enhanced T1w, and 2D T2w. In the context of our research, our focus centred exclusively on the utilisation of all the 53 3D contrast-enhanced T1w images.

The Montreal Neurological Institute’s Brain Images of Tumors (MNI BITE) database provides pre- and post-operative MR and ultrasound images obtained from patients diagnosed with brain tumours, as outlined in~\citep{mercierOnlineDatabaseClinical2012}. This collection encompasses 13 adult patients who were diagnosed with gliomas. These patients underwent both pre- and post-operative contrast-enhanced T1w MRI scans using the 1.5~T GE Signa EXCITE scanner. The two contrast-enhanced T1w imaging sessions (pre- and post-operative) were used for our study, resulting in 26 MRIs.

The demographic characteristics of these three databases are summarised in Table~\ref{tab:demographics}.

\begin{table}[tbh]
    \begin{center}
    \renewcommand{\arraystretch}{1.25}
    \caption{Distribution of the sex and age over the research (ADNI, MSSEG and MNI BITE) and the clinical (AP-HP) datasets used in our study.}
    \label{tab:demographics}
    \begin{tabular}{cccccc}
        \hline
        Database     & N°  subjects & N°  images  & Age {[}range{]}           & Sex (\%F) \\
        \hline
        ADNI         & 70         & 1143     &  74.31 ± 7.11  {[}55, 90{]}  & 41.43  \\
        MSSEG        & 53         & 53       & 45.42 ± 10.27 {[}24, 66{]} & 71.70  \\
        MNI BITE     & 13         & 26       & 52.00 ± 17.70 {[}31,76{]} & 35.71  \\
        \hline \hline
        AP-HP        & 3346       & 4045     & 55.15 ± 7.89 {[}18, 95{]} & 55.39 \\
        \hline
    \end{tabular}
    \end{center}
\end{table}

\subsection{Clinical dataset}
The routine clinical data come from an extensive CDW containing all the 3D T1w brain MRIs of adult patients scanned in various hospitals in the Greater Paris region associated with AP-HP. Given the large consortium of hospitals within the AP-HP (39 institutions) and the significant volume of images acquired on a daily basis, this CDW is a reliable and representative compilation of 3D T1w brain MRIs.

We used the same dataset as in our previous study, which was a random selection of 5500 images representing 4177 patients~\citep{bottaniAutomaticQualityControl2021}. These images were acquired using different scanner models from different manufacturers, including Siemens Healthineers (n = 3752, including 13 different scanner models), GE Healthcare (n = 1710, including 12 different scanner models), Philips (n = 33, including 3 different scanner models), and Toshiba (n = 5, including 2 different scanner models), as described in detail in ~\citep{bottaniAutomaticQualityControl2021}. All our 3D T1w MR images are based on gradient echo with inversion recovery, which corresponds to the Siemens MPRAGE, GE BRAVO and Philips TFE sequences. 
For simplicity, in the following we refer to the 3D T1w MR images based on gradient echo with inversion recovery as `3D T1w MRIs'.

\begin{figure}[ht]
   \centering
   \caption{Examples of manually labelled T1w MRIs of the CDW. For each type of artefact, we display an example of an image corrupted by a moderate and a severe artefact.}
   \includegraphics[width=0.9\linewidth]{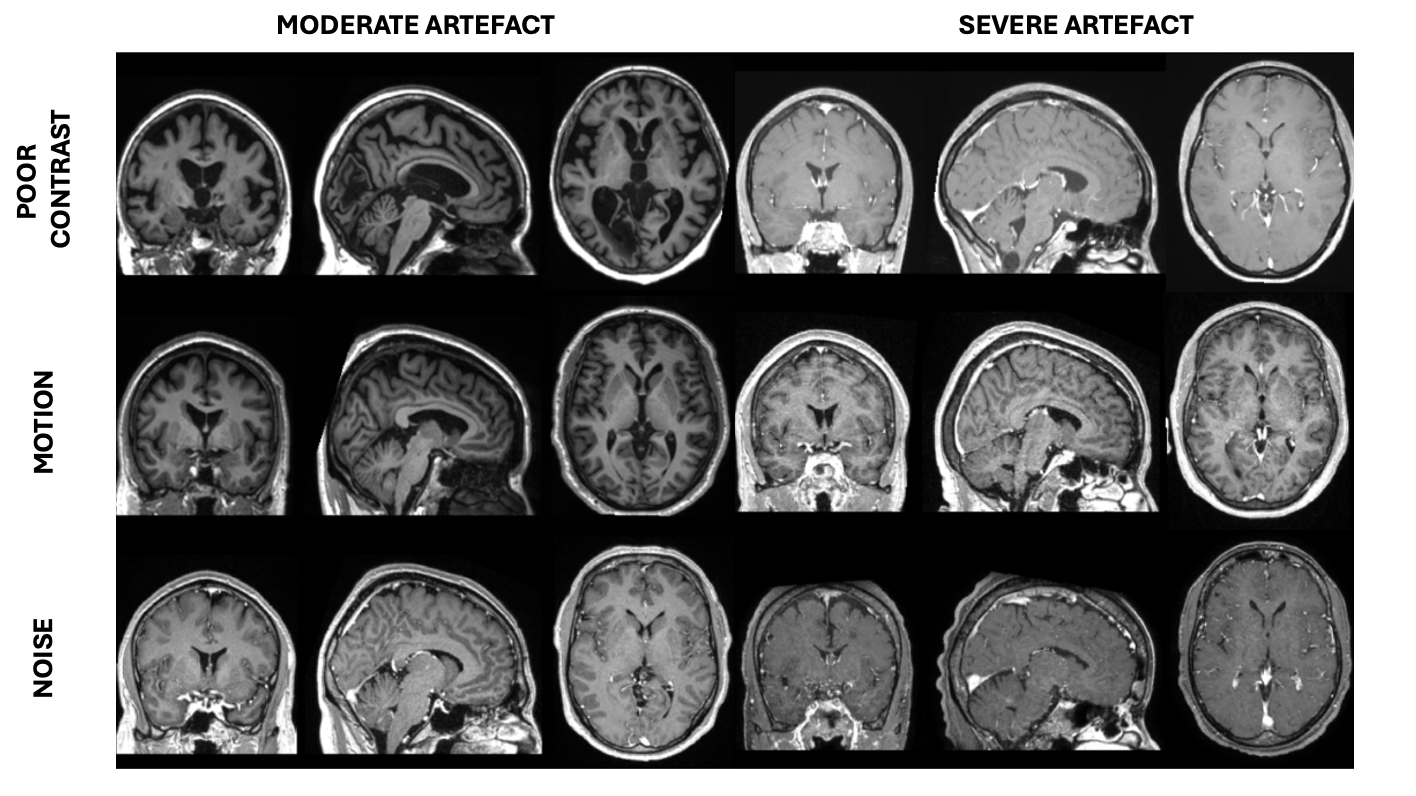}
   \label{fig:lab_samples}
\end{figure}

Manual annotations were previously performed by two raters regarding the quality of the 5500 T1w MRIs. Specifically, the contrast, noise and motion characteristics of the images were assessed using a three-level scale (Figure~\ref{fig:lab_samples}). The weighted Cohen's kappa between the two manual annotators for each of these artefacts is reported in the Appendix (Table~\ref{suptab:kappa}). Based on the grades attributed to these three characteristics, quality tiers were defined as follows: tier 1 stands for good quality MRI (grade of 0 for the three characteristics); tier 2 for medium quality (one characteristic with a grade of 1 and none with 2); tier 3 for bad quality (one characteristic with a grade of 2). The rules used to define the tiers are summarised in Table~\ref{tab:tier_description}. A total of 1455 images were categorised as ``straight reject'' because they were not true 3D T1w MRIs, mainly because they were truncated or segmented images, and were therefore excluded from this study. As a result, we obtained a dataset comprising 4045 images from 3346 patients. The demographic characteristics are shown in Table~\ref{tab:demographics}. From these MRIs, we randomly selected 385 images to build our independent test set, which is the same used in \citep{bottaniAutomaticQualityControl2021}. Notably, this test set had the same tier distribution as the images within the training and validation set.

\begin{table}[ht]
    \centering
    \caption{Rules used to define the quality control tiers.}
    \label{tab:tier_description}
    \renewcommand{\arraystretch}{1.25}
    \begin{tabular}{c|c}
        \hline
        Tier & Rule \\ 
        \hline \hline
        \makecell{Tier 1 (good quality)} & \makecell{Grade 0 for motion,  contrast and noise artefacts}   \\ 
        \hline
        \makecell{Tier 2 (medium quality)}  & \makecell{At least one type of artefact with grade 1; none with grade 2}   \\ 
        \hline
        \makecell{Tier 3 (bad quality)} & \makecell{At least one type of artefact with grade 2}\\ 
        \hline
    \end{tabular}
\end{table}

\section{Methods}
\begin{figure}
    \centering
    \caption{General workflow of the proposed transfer learning framework. First, artefact-free research MRIs are corrupted with synthetic artefacts (motion, noise and poor contrast). These images are used to pre-train artefact-specific CNNs (motion, noise and poor contrast). Models are then fine-tuned on routine clinical data relying on the manual annotation of these artefacts for 3660 MRIs. The overall quality tiers are then determined using the outputs of the artefact-specific models.}
    \includegraphics[width=0.9\linewidth]{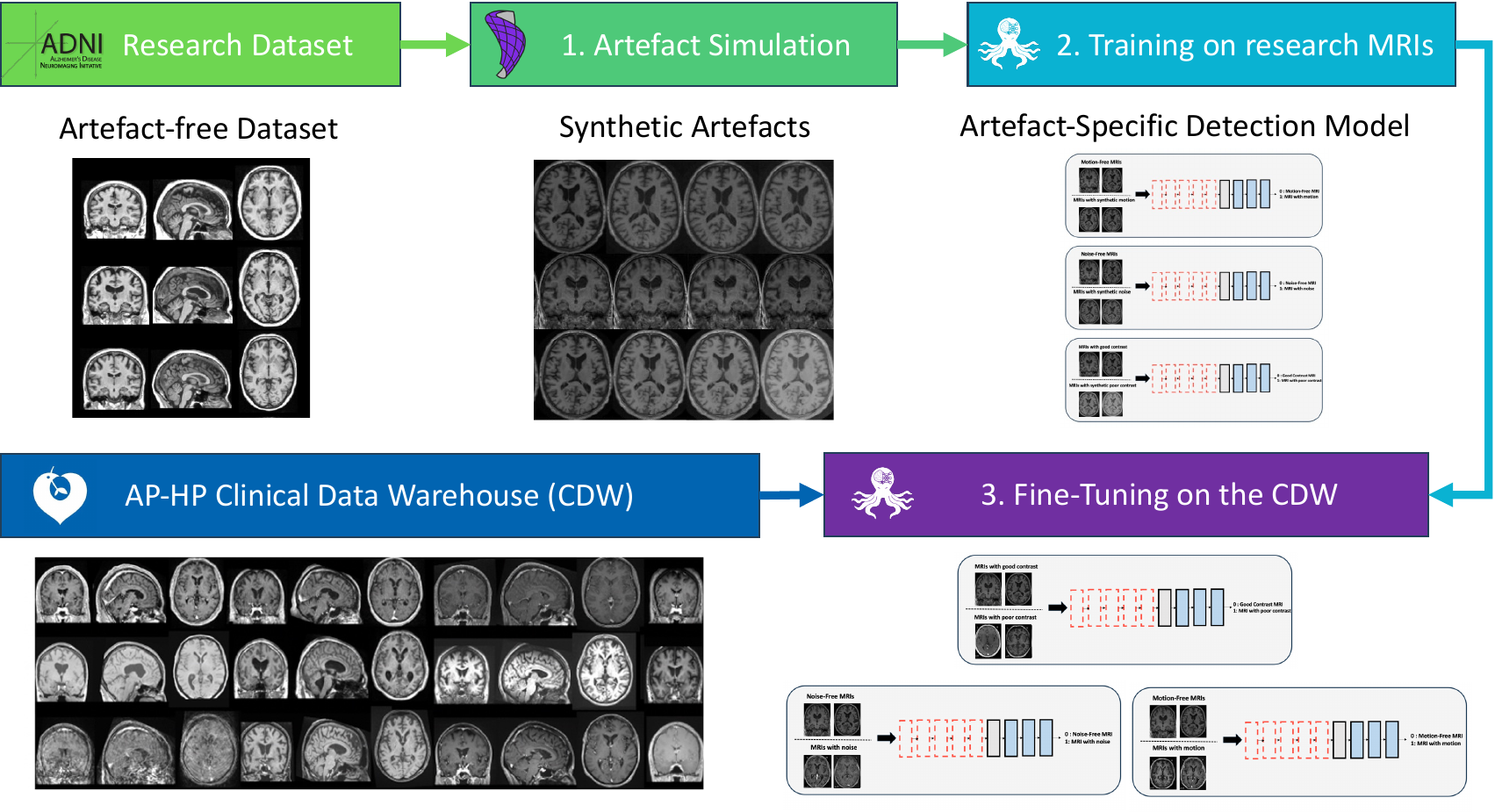}
    \label{fig:framework}
\end{figure}

In this paper, we have developed a new transfer learning approach to automate quality control in the CDW. Our proposed method is divided into three main steps. First, we pre-train three CNNs using images from the research-oriented dataset corrupted with synthetic artefacts to detect specific artefacts (motion, noise and poor contrast). Then, our three models are generalised to `real' artefacts on routine clinical 3D gradient echo T1w MRIs from the CDW, thanks to a transfer learning technique. Finally, we aggregate the results of these three fine-tuned models (noise, motion and contrast detection) to predict the overall image quality tier. The overall approach is shown in Figure~\ref{fig:framework}. We call this approach ``indirect quality tier classification'' (the tiers are not directly determined by a CNN but inferred from the results of three CNNs that each detect a specific type of artefact). This approach was compared to the 
``direct quality tier classification'', where the tiers are determined directly by a single CNN. In this case, the pre-training is performed using intentionally corrupted images with different types of artefacts. Each image is assigned a quality tier according to the severity of the generated synthetic artefacts. Figure~\ref{subfig:direct_indirect_workflow} illustrates the two distinct approaches.

\subsection{Artefact generation} 
\label{sec:art_gen}
The three types of artefact studied were simulated using the TorchIO library and its \texttt{RandomMotion}, \texttt{RandomGamma} and \texttt{RandomNoise} functions~\citep{perez-garciaTorchIOPythonLibrary2021}.  

\subsubsection*{Motion generation}
\begin{figure}[ht]
    \centering
    \includegraphics[width=1\linewidth]{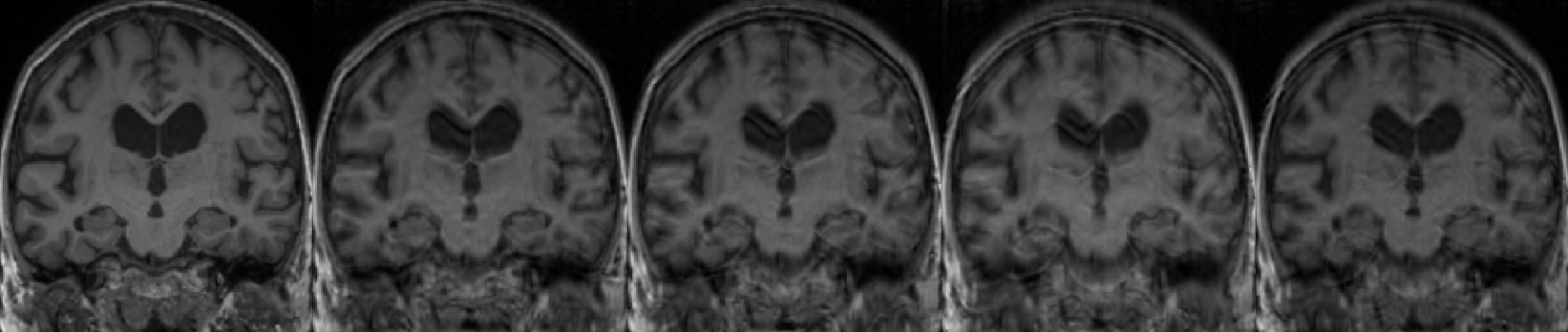}
    \caption{Simulated motion artefacts applied to an MRI of the ADNI dataset. From left to right: original MRI, rotation and translation of 2° and 2~mm, rotation and translation of 4° and 4~mm, rotation and translation of 6° and 6~mm, rotation and translation of 8° and 8~mm}
    \label{fig:adni_motion}
\end{figure}

We used the TorchIO function \texttt{RandomMotion} to simulate synthetic motion in T1w MR images from the research-oriented dataset (ADNI, MSSEG and MNI BITE) that already proved successful in our previous studies~\citep{Loizillon2023TransferLearning, loizillon2024automatic}. This function implements the simple image-based approach described in~\citep{shawMRIKSpaceMotion}, which assumes that the subject takes $Nt$ distinct positions during the acquisition. Head motion is approximated as a rigid body motion that comprises six degrees of freedom, three for the translation and three for the rotation~\citep{duffyRetrospectiveCorrectionMotion}. By playing with the rotation and translation ranges, we were able to simulate different motion artefact severity degrees (Figure~\ref{fig:adni_motion}). 

\subsubsection*{Noise generation}
\begin{figure}[ht]
    \centering
    \includegraphics[width=1\linewidth]{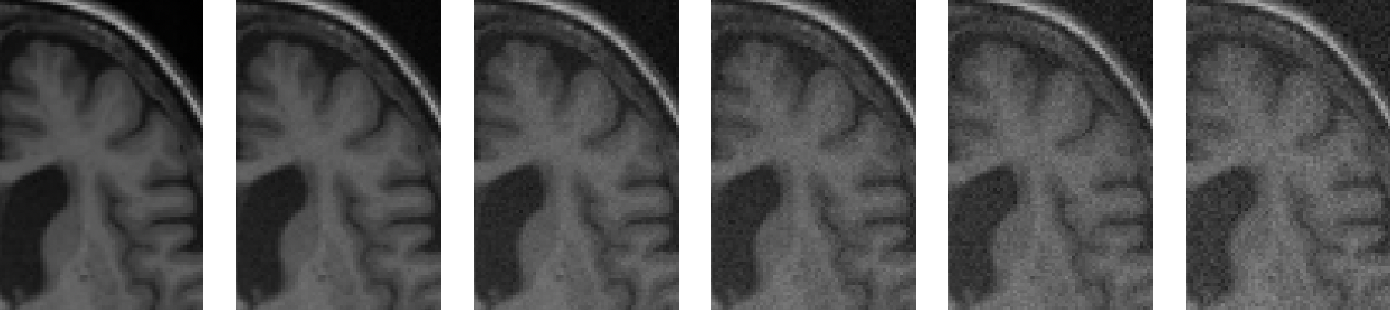}
    \caption{TorchIO \texttt{RandomNoise} function applied to an MRI of the ADNI dataset. From left to right: original MRI, $\sigma$=5, $\sigma$=10, $\sigma$=15, $\sigma$=20 and $\sigma$=25 }
    \label{fig:adni_noise}
\end{figure}

To mimic the presence of noise that can degrade 3D T1w MRIs, we artificially added synthetic noise to artefact-free images from our research-oriented dataset. The noise in an MRI typically follows a Rican distribution. For the sake of simplification, we adopt the common practice of considering the noise in the image domain as a Gaussian process with zero mean, no spatial Gaussian correlation, and equal variance in the real and imaginary parts~\citep{aja2016statistical}. To simulate Gaussian noise in MRIs, we employed the TorchIO function \texttt{RandomNoise}. Figure~\ref{fig:adni_noise} presents different severity degrees of simulated noise obtained by varying the standard deviation $\sigma$ of the Gaussian distribution from which the noise is sampled.


\subsubsection*{Contrast generation}
\begin{figure}[ht]
    \centering
    \includegraphics[width=1\linewidth]{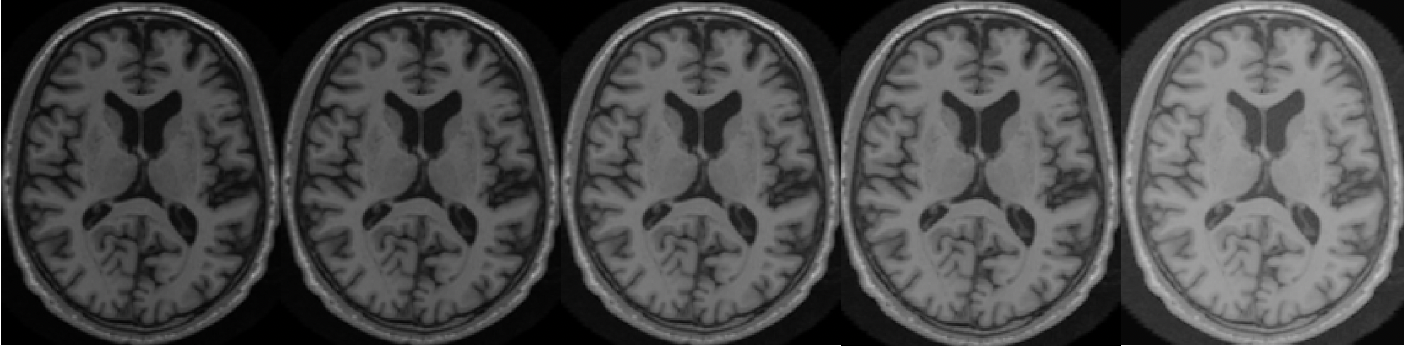}
    \caption{TorchIO \texttt{RandomGamma} function applied to an MRI of the ADNI dataset. From left to right: original MRI, $\beta$=-0.1, $\beta$=-0.4, $\beta$=-0.6, $\beta$=-0.8 }
    \label{fig:random_gamma}
\end{figure}

In T1w MRI, the quality of the contrast is judged by the ability to distinguish between white matter and grey matter. Using TorchIO's \texttt{RandomGamma} function, we apply a non-linear gamma correction to simulate poor contrast in good quality MRIs. For an artefact-free MR image denoted as $I$,  we generate poor contrast via the parameter $\beta$ of the gamma correction function to obtain our new synthetically corrupted MRI denoted as $I_{\text{c}}(\beta) = I^{1/{e^\beta}}$. In Figure~\ref{fig:random_gamma}, various levels of simulated poor contrast are illustrated, achieved by adjusting the $\beta$ parameter within the \texttt{RandomGamma} function.

\subsection{Artefact quantification} 

Before pre-training the artefact-specific models, we ensured that realistic motion, noise and poor contrast artefacts were simulated in the three research databases: ANDI, MSSEG and MNI BITE. We used clinical routine images to determine appropriate parameter values for simulating various types of artefacts in the research-oriented dataset, aiming to replicate the moderate and severe artefact levels observed in the CDW. Initially, we quantified the severity of artefacts present in images of the CDW for both moderate and severe cases. Subsequently, we identified the relevant parameters to simulate moderate and severe artefacts in the research dataset, ensuring they corresponded to those observed in the clinical routine dataset.

To quantify the contrast of a given MRI, $I$, where $\bar{I}_{\text{WM}}$ and $\bar{I}_{\text{GM}}$ are the mean intensities of respectively white and grey matter voxels, we defined the normalised difference of white and grey matter brain tissue (ND-WGM) as follows:

\begin{equation}
\text{ND-WGM} = \left|\frac{{\bar{I}_{\text{WM}} - \bar{I}_{\text{GM}}}}{{\bar{I}_{\text{WM}} + \bar{I}_{\text{GM}}}}\right| \enspace .
\end{equation}
We perform the segmentation of the grey and white matter tissues thanks to the FSL tools \texttt{BET} and \texttt{FAST}~\citep{jenkinson2012fsl} to compute the ND-WGM for a randomly selected set of 50 routine clinical T1w MRIs for each level of contrast severity (\textit{cont\_0}, \textit{cont\_1} and \textit{cont\_2}). This selection of 150 MRIs accurately reflects the diversity of contrast severity levels within our clinical dataset. We ensure that the segmentation performed by \texttt{FAST} was correct by manually inspecting them to exclude images with segmentation errors.

The quantification of realistic noise in T1w MRIs relies on the signal to noise ratio (SNR). We denote $I$ as the T1w MRI and defined $\bar{I}_{\text{WM}}$ as the mean intensity of white matter voxels of $I$ and $\sigma_{\text{AIR}}$ as the standard deviation of the air in $I$. The SNR is computed as
\begin{equation}
\text{SNR} = \frac{{I_{\text{WM}}}}{\sigma_{\text{AIR}}} \enspace .
\end{equation}
We evaluate the SNR for 50 manually annotated MRIs from the CDW per noise severity (except for score 2, for which we have only 16 MRIs) that were randomly chosen and which reflect the wide spectrum of noise severity present in this dataset. Similar to contrast quantification, we rely on \texttt{BET} and \texttt{FAST} to obtain the segmentation masks, which have been manually inspected to reject segmentation errors, allowing us to calculate the SNR. 

The evaluation of these two metrics on the routine clinical data allows us to determine the appropriate $\sigma$ and $\beta$ parameters of the \texttt{RandomNoise} and \texttt{RandomGamma} functions for the simulation of moderate and severe artefacts on the research data to pre-train our models. For noise simulation, we explored various standard deviation ($\sigma$) ranges to find optimal parameters for corrupting MRIs with moderate and severe noise: $\sigma$ = \{[0, 10]; [5, 15]; [10, 20]; [15, 25]; [20, 30]; [25, 35]\}. For the simulation of poor contrast, we investigated the following $\beta$ parameter ranges: $\beta$ = \{[-0.2, -0.05]; [-0.25, -0.15]; [-0.30, -0.15]; [-0.35, -0.20]; [-0.40, -0.25]; [-0.45, -0.03]; [-0.50, -0.35]\}. We retained the two ranges of $\sigma$ and $\beta$ leading to the mean value of SNR and ND-WGM closest to those obtained on clinical routine images for moderate and severe artefacts.

Regarding motion artefacts, there is still no robust metric in the literature to quantify the motion present in MRIs. Recently, ~\cite{eichhorn2022evaluating} have shown that SSIM and PSNR were the metrics that correlate best with radiological assessment. What's more, they relied on the evaluation of pixel-by-pixel differences between images and were therefore very sensitive to misregistration between the original image and the corrupted image~\citep{reguig2022global}. These metrics also have the disadvantage of requiring a reference image, making it impossible to quantify motion artefacts within our manually annotated clinical routine images. \cite{eichhorn2022evaluating} also propose two different reference-free metrics for motion quantification (average edge strength and Tenengrad measure). However, due to their sensitivity to many cofactors such as
contrast, neither of these metrics showed a significant correlation with our manual annotations (Figures~\ref{supfig:motion_compute}, \ref{supfig:aes_contrast} and \ref{supfig:tgn_contrast}). Thus, the parameters found in our previous study were re-used to generate different levels of motion~\citep{Loizillon2023TransferLearning, loizillon2024automatic}.

\subsection{Network pre-training on research data}

We aimed to first detect specific artefacts (motion, noise and poor contrast) in the research-oriented dataset thanks to three independent networks trained with simulated artefacts. For each type of artefact, we have two tasks: the detection of moderate and severe artefacts. Thus, six different models were trained in total as we trained our two tasks on three different types of artefact (motion, noise and poor contrast).

We defined an independent test set by randomly selecting 61 artefact-free images over the research-oriented dataset and corrupting them with moderate to severe noise, motion and poor contrast artefacts. The remaining 1161 good quality images were also corrupted with synthetic artefacts and were split into training and validation using a 5-fold cross validation (CV). More details about the train/validation/test splits for the different experiments are given in the Supplementary Materials (Table~\ref{subtab:distribution_research} and~\ref{subtab:distribution_research_severe}). To achieve robustness in artefact detection, our models were trained to handle cases where the label 0 for a specific artefact (e.g., ``Mov0") includes MRIs that are corrupted with other types of artefacts (e.g., poor contrast or noise). For instance, in Table~\ref{subtab:distribution_research}, the 2859 training images labelled with ``Mov0" consist of 953 artefact-free MRIs, 953 images corrupted with moderate contrast and 953 images with moderate noise. 

For each experiment, among the five CV models, the one with the lowest loss on the validation set was saved as the final pre-trained model.

\subsection{Network fine-tuning on routine clinical data}

To generalise our artefact-specific networks to clinical routine data, we fine-tuned the six pre-trained models on the 3660 manually labelled MRIs from the AP-HP CDW on six specific target tasks: detecting severe artefacts (Contrast01vs2, Noise0vs12, Motion01vs2) and moderate ones (Contrast0vs1, Noise0vs1, Motion0vs1). Models were evaluated on an independent test set of 385 routine clinical MRIs. Note that for noise, we perform the task Noise0vs12 due to the limited images labelled with severe noise (noise2: 16 MRIs). This modification allows us to effectively leverage the images available and create a more balanced dataset. The detailed distribution of the training, validation and test sets for our fine-tuned models can be found in the Supplementary Materials (Table~\ref{subtab:distribution_clinical_severe} and Table~\ref{subtab:distribution_clinical_moderate}).

The inference step is divided into two steps: (1) the evaluation of the performance on each artefact-specific network; (2) the re-combination of the overall quality tier based on the three grades corresponding to the score for motion, noise and contrast. 


In each experiment, the final model was selected from the five CV models based on the lowest loss over the validation set.

\subsection{Implementation details}
The architecture of the CNNs we used corresponds to that of the 3D Conv5FC3 network which had proven its effectiveness in our previous study~\citep{bottaniAutomaticQualityControl2021}. The Conv5FC3 is composed of five convolutional blocks -- comprising a convolutional layer, a batch normalisation layer, a ReLU activation function and a maximum pooling layer -- and three fully connected layers. For the hyperparameters, the Adam optimiser was chosen, setting the learning rate to 1e-4 and the batch size to 6. The weighted binary cross entropy was employed as the loss function. Final models were chosen as being the ones with the lowest loss over the validation set. Fine-tuning on the clinical routine data was then applied by retraining only the last three fully connected layers in order to generalise our pre-trained models to `real' artefacts from routine clinical data. Thus, we were able to reduce the gap between research data with synthetic artefacts and clinical data with real artefacts. 
The Conv5FC3 architecture, along with the fine-tuning process and artefact simulation, has been fully implemented and is accessible within the ClinicaDL software repository on GitHub (\url{https://github.com/aramis-lab/clinicaDL})~\citep{thibeau-sutreClinicaDLOpensourceDeep2021}.

\section{Results}

\subsection{Artefact simulation}

Through various tests of the $\sigma$ and $\beta$ values to generate noise artefacts and different contrast levels, we were able to identify the most relevant parameters in terms of ND-WGM and SNR that were consistent with those obtained in the CDW. Parameters used for generating moderate and severe artefacts in the research-oriented dataset using the \texttt{RandomMotion}; \texttt{RandomNoise} and \texttt{RandomGamma} are summarised in Table~\ref{tab:torchio_param}. 

\begin{table}[ht]
    \caption{TorchIO parameters for the artefact simulation of moderate and severe noise, motion and poor contrast}
    \label{tab:torchio_param}
    \centering
    \begin{small}
    \begin{tabular}{|c|c|c|c|}
        \hline 
        & \texttt{\makecell{Random  \\Motion}} & \texttt{\makecell{Random  \\Gamma}} 
          & \texttt{\makecell{Random  \\Noise}} \\
        \hline \hline
        Moderate & \makecell{R= {[}2°, 4°{]}; \\ T= {[}2 mm, 4 mm{]}} & $\beta$ = [-0.2, -0.05] & $\sigma$ =[5, 15]  \\
        \hline
        Severe & \makecell{R: {[}5°, 8°{]}; \\ T: {[}5 mm, 8 mm{]}}  &  $\beta$ = [-0.45, -0.3] & $\sigma$ =[15, 25] \\
        \hline
    \end{tabular}    
    \end{small}
\end{table}

\begin{figure}[ht]
    \centering
    \caption{Violin plot of the distribution of normalised difference of white and grey matter (ND-WGM) and signal to noise ratio (SNR) for the real images from the CDW (blue) and the images with synthetic artefacts from the ADNI research dataset (orange). The median is represented by a dotted line, while the interquartile ranges are indicated by narrower dotted lines.} 
    \includegraphics[width=\linewidth]{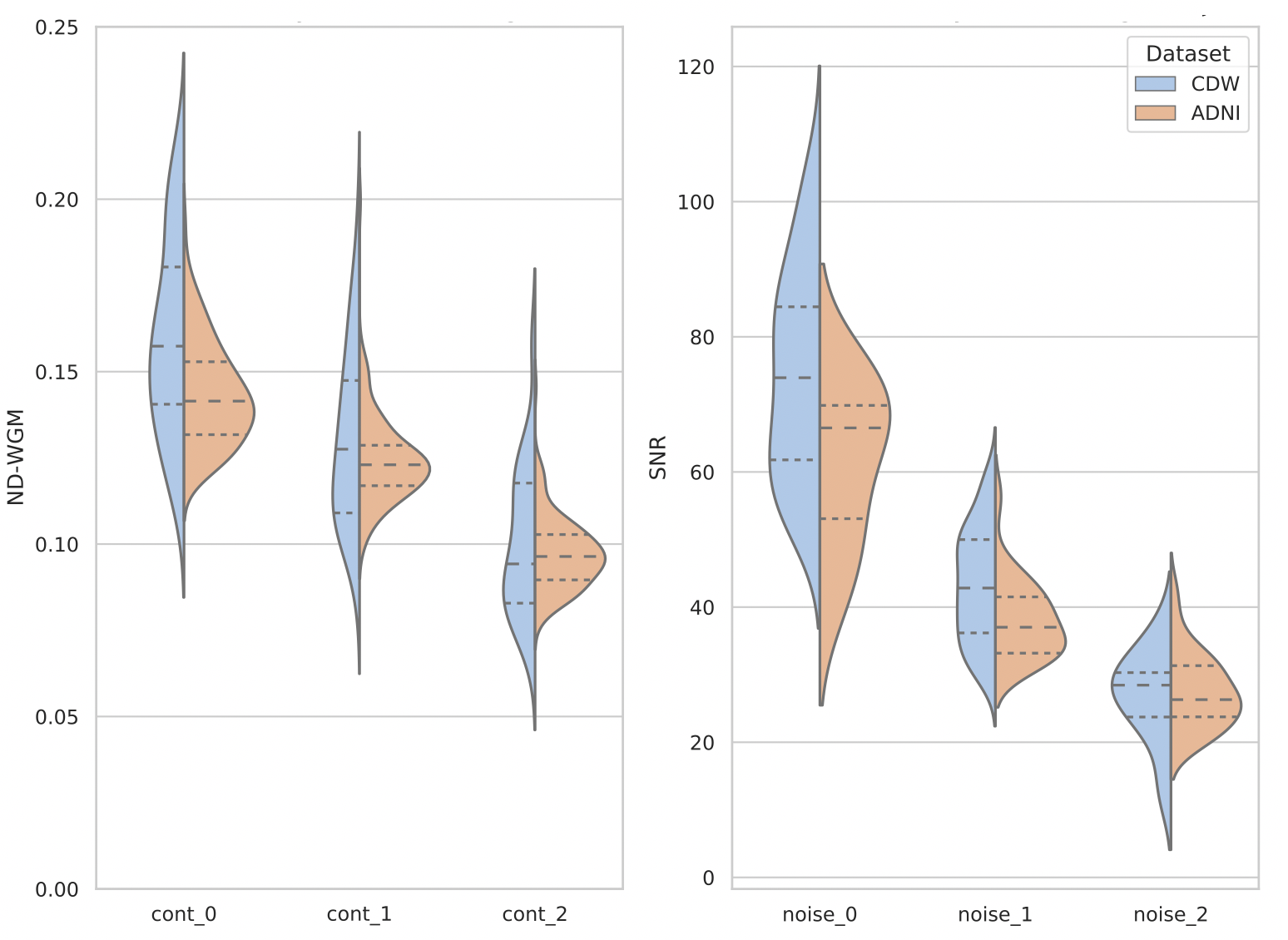}
    \label{fig:SNR_ND_CDW_ADNI}
\end{figure}

Figure~\ref{fig:SNR_ND_CDW_ADNI} illustrates the distribution of the SNR and ND-WGM through a violin plot, depicting a subset of the clinical routine dataset and of the ADNI dataset with synthetic artefacts. 

    For contrast quantification, we present the distribution of 48 cont\_0, 48 cont\_1 and 38 cont\_2 routine clinical MRIs of the CDW. Indeed, our manual inspection of the \texttt{FAST} results showed that the segmentation failed on 2, 2 and 12 images with scores of 0, 1 and 2 respectively. The ND-WGM results present different means depending on the contrast levels manually assigned to the MR images. For T1w MRIs with a score of 0, indicating good contrast, the average ND-WGM was 0.16. For those with a score of 1, indicating moderately poor contrast, the average was 0.13. Finally, MRIs with a poor contrast score of 2 have an average ND-WGM of 0.10. These results show consistency: the lower the contrast score assigned by the manual annotators, the lower the ND-WGM, highlighting the direct impact of contrast quality on ND-WGM.
    We synthesised two levels of contrast, where $\beta$ values were selected to ensure that the ND-WGM measures of corrupted MRIs closely matched the one of \textit{cont\_1} and \textit{cont\_2}. To simulate moderate contrast, corresponding to \textit{cont\_1}, we used a range of $\beta$ values from -0.2 to -0.05. For obtaining poor contrast, to mimic \textit{cont\_2} images, we used $\beta$ in the range [-0.45,-0.3].
    A convincing distribution alignment regarding the ND-WGM was obtained between the research-oriented dataset with synthetic artefacts and the CDW dataset. For ADNI MRIs corrupted by moderate contrast, we measured a mean ND-WGM of 0.124, compared with 0.131 in routine clinical data. Very poor contrast added into ADNI MRIs showed a mean of 0.097, against 0.101 in clinical data.

    For noise quantification, we plot the distribution for 45 noise\_0 labelled MRIs, 48 noise\_1 MRIs and only 9 MRIs with the noise\_2 label, as our examination of the \texttt{FAST} results revealed that segmentation failures occurred in 5, 2 and 7 images with a noise score of 0, 1 and 2. On average, MRIs labelled as noise-free (score 0) had a SNR of 74, while those labelled as moderately noisy (score 1) had a SNR of 44. Finally, MRIs presenting severe noise (score of 2) had an average SNR of only 25. Similar to the results obtained for contrast with ND-WGM, these results consistently show that the worse the score assigned by the human annotators, the lower the SNR of the MRI.
    To align the SNR distribution of the ADNI dataset with the routine clinical MRIs labelled with \textit{noise\_1} and \textit{noise\_2}, we generate moderate noise using a standard deviation range of $\sigma \in [5,15]$ and severe noise was added using a range of $\sigma \in [15,25]$. 
    By simulating noise in artefact-free research images, we harmonised the SNR distribution with that of routine clinical data. In particular, our noise simulation yielded an average SNR value of 27 for severe noise, a value that matches the one of noise\_2 MRIs of the CDW (SNR=27). Similarly, when we simulated moderate noise within the ADNI research dataset, we achieved an average SNR of 38, which aligns the 44 SNR obtained from the clinical routine data labelled with noise\_1. 

These distribution alignments for both noise and contrast metrics highlight our ability to replicate the artefact characteristics of clinical data within research datasets. For the motion simulation, due to the lack of reliable metrics to quantify motion artefacts in MRI, we used the same parameters that proved successful in our previous studies~\citep{Loizillon2023TransferLearning, loizillon2024automatic}. Thus, we assumed that the subject takes $Nt=4$ distinct positions during the acquisition and moderate and severe motion were generated using the following parameters: rotation [2°, 4°], translation [2~mm, 4~mm] and rotation [5°, 8°], translation: [5~mm, 8~mm], respectively. 

\subsection{Pre-training performance on research dataset}
The ability of deep learning models to detect each type of artefacts (motion, noise and poor contrast) was first assessed using images from the research-oriented dataset corrupted with synthetic artefacts. We evaluated the performance of our Conv5FC3 model trained on synthetic artefacts when applied to our synthetic independent test set corrupted with different artefact severity degrees. The balanced accuracy on our independent test set is excellent for severe artefact detection (motion: 100~\%, noise: 97.78~\%, poor contrast: 97.57~\%) and very good to excellent for moderate artefacts detection (motion: 99.54~\%, noise: 87.35~\%, poor contrast: 90.17~\%).

We also assessed the performance of our model directly trained for the tier classification using synthetic tiers on research dataset. On our independent synthetic test set, we obtained a balanced accuracy of 92.04~\% for the Tier 1-2 vs. Tier 3 task and 91.00~\% for the Tier 1 vs. Tier 2 task.

\subsection{Automatic tier classification on routine clinical data}

Results obtained for our two classification tasks, the detection of bad quality MRIs (Tier 1-2 vs. Tier 3) and moderate quality MRIs (Tier 1 vs. Tier 2) with our proposed transfer learning technique are presented in Table~\ref{tab:results_tier_2}.  
We report the comparison of our proposed approach  with the direct tier classification, where we directly pre-trained a single model using intentionally corrupted images with different types of artefacts, which are labelled with a quality tier according to the severity of the generated synthetic artefacts (Figure~\ref{subfig:direct_indirect_workflow}). For both approaches, namely the indirect and direct tier classification, we explore the performance of models trained from scratch or fine-tuned on the CDW. We also report the balanced accuracy of the manual annotators, which is defined as the average of the balanced accuracy between each rater and the consensus label.

For the detection of bad quality images (Tier 1-2 vs. Tier 3), the training/validation set includes 3660 images; and for the detection of moderate quality images (Tier 1 vs. Tier 2), the training/validation set includes 2182 images.

Our proposed indirect tier classification achieved a good balanced accuracy of 87.05\% for the Tier 1-2 vs. Tier 3 task, which is 3.48 percent points higher than our previous SOTA approach based on the direct tier classification trained from scratch  (\cite{bottaniAutomaticQualityControl2021}: 83.57\%), but lower than that of the manual annotators (91.56\%). Same trends were observed for the most difficult task (Tier 1 vs. Tier 2), the classifier is good (79\%) and outperforms the classifier trained from scratch using direct tier classification by 4.95 percent points, but its performance is lower than that of the manual annotators (88.27\%). Some failure cases are shown in the Appendix (Figure~\ref{supfig:failure_cases}).

The transfer learning approach, utilising synthetic artefacts on research datasets, outperformed the training from scratch for both direct and indirect techniques and regardless of task (Tier 1 vs. Tier 2  and Tier 1-2 vs. Tier 3).
Compared to the direct tier classification results presented in our previous study~\citep{bottaniAutomaticQualityControl2021}, the indirect approaches yielded better performance both when training from scratch and when fine-tuning the models on the CDW. We gain more than 3.5 percent points on the Tier 1 vs. Tier 2 task and more than 2 percent points on the Tier 1-2 vs. Tier 3 task over the fine-tuning approach using direct tier classification.

\begin{table}[ht]
    \centering
    \caption{ Balanced accuracy (b. acc.) of the indirect and direct classification approaches for the two tasks: Tier 1 vs. Tier 2 and Tier 1-2 vs. Tier 3. For both approaches and both tasks, we report the performance obtained from scratch (training from scratch on the CDW) and by fine-tuning on the clinical routine data (fine-tuning on the CDW). We also report the agreement between human raters and the consensus (b. acc. annotators). Results with ** indicate a statistically significant difference (corrected p$<$0.05, Wilcoxon signed rank test) with respect to the proposed approach (indirect tier classification using fine tuning on the CDW).}
    \label{tab:results_tier_2}
    \begin{tabular}{lccc}
    \hline
        & & Tier 1 vs Tier 2            & Tier 1-2 vs  Tier 3 \\
        \hline \hline
        & B. acc. annotators                   & 88.27                & 91.56                \\ 
        \hline
        \multirow{2}{*}{\makecell{Indirect Tier\\   Classification}  }& \makecell{Training from scratch\\   on the CDW}   & 76.11                &    84.83$^{**}$                  \\ \cline{2-4}
        & \makecell{Fine-tuning on the CDW\\  (proposed)}  & 79.00                &    87.05                  \\ 
        \hline
        \hline 
        \multirow{2}{*}{\makecell{Direct Tier \\ Classification}  } & \makecell{Training from scratch\\  on the CDW \\ ~\citep{bottaniAutomaticQualityControl2021}}   & 74.05$^{**}$         & 83.57         \\  \cline{2-4}
        &  Fine-tuning on the CDW & 75.41$^{**}$        & 84.91         \\ 
        \hline
    \end{tabular}
\end{table}

As the indirect tier classification is based on the recombination of the output of three artefact-specific models, we analyse in more detail the performance of each of these models. Table~\ref{tab:results_artefact_specific} summarises the balanced accuracy obtained by the three artefact-specific models for both the detection of moderate and severe artefacts. Fine-tuning improved the performance of all six models, bringing them closer to the balanced accuracy of manual annotators. For the severe artefact detection tasks, the models fine-tuned on routine clinical data outperformed the ones trained from scratch by more than 1, 5 and 8 percent points for noise, contrast and motion detection, respectively. The same trend was observed for the moderate detection task, where fine-tuning resulted in a gain of 6 percent points for both motion artefact and poor contrast detection.

\begin{table}[ht]
    \centering
    \caption{Detection of motion, noise and poor contrast artefacts on the CDW. For both the detection of severe and moderate artefacts, we report: the agreement between human raters and the consensus (b. acc. annotators), results of the proposed fine-tuning approach (fine-tuning on CDW) and results when training from scratch on CDW (training from scratch on CDW). Severe: severe artefact detection; Moderate: moderate artefact detection; b. acc.: balanced accuracy.}
    \label{tab:results_artefact_specific}
    \renewcommand{\arraystretch}{1.25}
    \begin{small}
        \begin{tabular}{lcccc}
            \hline
                                            & & Motion            & Noise & Contrast            \\ 
            \hline
            \multirow{3}{*}{\begin{turn}{90} \small Moderate \end{turn}} & B. acc. annotators                   & 73.21                & 87.22 & 84.87                \\ 
            &Training from scratch on CDW    &   58.31      & 84.65 &    67.86    \\ 
            & Fine-tuning on CDW    & 64.35        & 88.39 & 74.06        \\
            \hline \hline
            \multirow{3}{*}{\begin{turn}{90} \small Severe \end{turn}} & B. acc. annotators                   & 86.24                & 91.03 & 87.92                \\ 
            &Training from scratch on CDW   &  73.75  & 85.97  & 84.93 \\ 
            &Fine-tuning on CDW    &   82.47     & 87.44 & 90.58        \\ 
            \hline
        \end{tabular}
    \end{small}
\end{table}

\section{Discussion}

    In this study, we implemented a transfer learning framework that bridges the gap between research and clinical MRIs for automating the quality control of 3D T1w brain MRIs in a CDW. To achieve this, we pre-trained artefact-specific CNNs on research data corrupted by realistic synthetic artefacts to detect images affected by poor contrast, motion and noise. We then used a transfer learning technique to adapt our pre-trained models to clinical data. Finally, the overall image quality is inferred from the results of CNNs, each designed to detect a specific type of artefact. The effectiveness of our approach was validated on an independent dataset of 385 manually annotated routine clinical MRI scans from 39 different hospitals affiliated to the AP-HP, all gathered in a CDW. We achieved excellent results in detecting poor quality MRIs with a balanced accuracy over 87\% and a satisfactory balanced accuracy of over 79\% in detecting moderate quality MRIs.
    
    
    For the detection of severe artefacts, our three fine-tuned models achieved excellent results with a balanced accuracy over 82~\%, 87~\% and 90~\% for the motion, noise and poor contrast artefacts, respectively. It is worth noting that the use of our transfer learning approach utilising synthetic artefacts on research datasets significantly improved our results, by up to 9 percent points compared to training the model from scratch using clinical data. This translates to an improvement from 73.75\% to 82.47\% for motion detection, from 85.97\% to 87.44\% for noise detection, and from 84.93\% to 90.58\% for poor contrast detection.
    For the most difficult task of detecting moderate artefacts, the fine-tuning strategy enables improving each artefact detection from 3.5 to 6 percent points compared to models trained from scratch on the CDW.

    The artefact-specific networks enable us to determine the quality tier of a given image. By doing so, we achieved a satisfying balanced accuracy of respectively 79.00~\% and 87.05~\% for the detection of moderate quality MRIs (Tier 1 vs. Tier 2) and the detection of bad quality MRIs (Tier 1-2 vs. Tier 3) tasks. Our proposed indirect tier classification approach surpasses the performance of our previous SOTA method~\cite{bottaniAutomaticQualityControl2021}, which relied on direct tier classification with training from scratch. There was a notable improvement of 4.95 and 3.48 percentage points for Tier 1 vs. Tier 2 and Tier 1-2 vs. Tier 3, respectively. Furthermore, indirect tier classification has added an important layer of information for understanding why an image is rejected. This approach enriches our model, not only by improving its predictions for both tasks, but also by providing important information about the underlying reasons for rejection.
    In the scope of our work, we also compared our transfer learning approach with training on the clinical routine data from scratch. The transfer learning strategy relying on the synthesis of artefacts on research data has proven beneficial in our experiments, whether for direct or indirect tier classification, improving the balanced accuracy up to 3.5 percent points.
    
    In our study, we put particular effort into generating realistic artefacts to corrupt the research datasets. We quantified the level of noise and contrast within the CDW MRIs thanks to the two following metrics: the signal to noise ratio and the normalised difference between grey and white matter. This allowed us to effectively choose the parameters $\beta$ and $\gamma$ of the TorchIO function \texttt{RandomNoise} and  \texttt{RandomGamma} to corrupt research data and align the distribution of SNR and ND-WGM with the one of routine clinical data.
    One of the limitations of our work in aligning real artefacts from routine clinical images with simulated ones in images from research datasets is that it could only be performed on the ADNI dataset. This is due to the need for tissue segmentation using \texttt{FAST} to compute ND-WGM and SNR, which is not possible on injected MR images. Consequently, our good results for the alignment of real and synthetic artefacts must be tempered because their generalisation to injected images could not be verified on the MSSEG and MNI BITE datasets. 
    What's more, when it came to motion, we had to rely on previously used parameters~\citep{Loizillon2023TransferLearning, loizillon2024automatic} since we could not find a reliable metric that did not require a reference image to quantify the amount of motion in a given MRI. Despite the recent efforts led by \cite{eichhorn2022evaluating} to find metrics that best correlate with radiological assessment, such as the average edge strength and Tenengrad measure, so far none of them have proven their robustness to other types of artefacts and in particular different types of contrast severity, leaving motion quantification as an open question.
    
    The evaluation of SNR and ND-WGM confirmed the crucial need for an automated QC tool within the CDW. While most good quality, non-injected images (tier1) were correctly segmented by \texttt{FAST}, the success of segmentation decreased as image quality deteriorated (24\% of failures on MRIs with severe contrast and 44\% on MRIs affected by severe noise). This is highlighting the impossible use of tools such as MRIQC~\citep{estebanMRIQCAdvancingAutomatic2017}, which relies mainly on the calculation of metrics extracted from segmentation maps, in a clinical routine context. 
    It is also important to acknowledge that our modelling of the different artefacts is simplified. We represent head motion as a rigid body motion added in the image domain which is randomly sampled from probability density functions, rather than in the k-space domain. Similarly, noise is modelled in a simplified manner as a Gaussian process with equal variance in both the real and imaginary domains. We use a non-linear gamma correction function to create different levels of bad contrast, but other approaches such as smoothing the images would also have been a way of synthetically corrupting the contrast of the image. It is worth noting, however, that these simplified models effectively serve our specific use case, which is model pre-training. Furthermore, in a previous study~\citep{loizillon2024automatic}, we demonstrated that the use of more complex motion artefact modelling approaches, such as k-space based methods, did not lead to significant improvements in motion artefact detection.
    
    Finally, our proposed method is currently only able to transfer the knowledge of the pre-trained models learned on gradient echo research MRIs with synthetic artefacts to routine clinical gradient echo MRIs with real artefacts. The generalisation of our models to other new types of sequences that are starting to become common in clinical routine, such as 3D spin echo or 3D FLAIR MRI, is of great interest. This is left for future work.

\section{Conclusion} 
In this study, we introduced an innovative transfer learning framework that uses realistic artefact simulation to facilitate automatic quality control of 3D gradient echo T1w brain MRIs within a large clinical data warehouse. Our approach involved the generation of poor contrast, motion and noise artefacts on research datasets, mimicking what is observed in the CDW. Based on these simulations, we pre-trained three artefact-specific CNNs, subsequently generalising them to clinical routine images through transfer learning, leveraging the labelling of a large dataset of 3660 MRIs.  Finally, the quality tiers were inferred from the results of the three CNNs that each detect a specific type of artefact.
Our deep learning classifiers showed a good ability to identify medium and poor quality images. This research highlights the importance of synthetic artefact generation and transfer learning for improving automated MRI quality assessment in routine clinical data. Our results contribute to the advancement of quality control procedures in routine clinical MRI by improving the reliability and efficiency of image quality sorting, which is essential in clinical data warehouses.


\acks{The research was done using the Clinical Data Warehouse of the Greater Paris University Hospitals. The authors are grateful to the members of the AP-HP WIND and URC teams, and in particular Yannick Jacob, Julien Dubiel, Antoine Roz\`{e}s, Cyrina Saussol, Rafael Gozlan, St\'{e}phane Br\'{e}ant, Florence Tubach, Jacques Ropers, Christel Daniel, and Martin Hilka. They would also like to thank the ``Coll\'{e}giale de Radiologie of AP-HP'' as well as, more generally, all the radiology departments from AP-HP hospitals.
The authors would also like to thank Romain Valabregue and Ghiles Reguig for their help implementing the k-space motion simulation and their feedback.

The research leading to these results has received funding from the Abeona Foundation (project Brain@Scale), from the French government under management of Agence Nationale de la Recherche as part of the ``Investissements d'avenir'' program, reference ANR-19-P3IA-0001 (PRAIRIE 3IA Institute) and reference ANR-10-IAIHU-06 (Agence Nationale de la Recherche-10-IA Institut Hospitalo-Universitaire-6). 

Data collection and sharing for this project was funded by the Alzheimer's Disease Neuroimaging Initiative (ADNI) (National Institutes of Health Grant U01 AG024904) and DOD ADNI (Department of Defense award number W81XWH-12-2-0012). ADNI is funded by the National Institute on Aging, the National Institute of Biomedical Imaging and Bioengineering, and through generous contributions from the following: AbbVie, Alzheimer’s Association; Alzheimer’s Drug Discovery Foundation; Araclon Biotech; BioClinica, Inc.; Biogen; Bristol-Myers Squibb Company; CereSpir, Inc.; Cogstate; Eisai Inc.; Elan Pharmaceuticals, Inc.; Eli Lilly and Company; EuroImmun; F. Hoffmann-La Roche Ltd and its affiliated company Genentech, Inc.; Fujirebio; GE Healthcare; IXICO Ltd.; Janssen Alzheimer Immunotherapy Research \& Development, LLC.; Johnson \& Johnson Pharmaceutical Research \& Development LLC.; Lumosity; Lundbeck; Merck \& Co., Inc.; Meso Scale Diagnostics, LLC.; NeuroRx Research; Neurotrack Technologies; Novartis Pharmaceuticals Corporation; Pfizer Inc.; Piramal Imaging; Servier; Takeda Pharmaceutical Company; and Transition Therapeutics. The Canadian Institutes of Health Research is providing funds to support ADNI clinical sites in Canada. Private sector contributions are facilitated by the Foundation for the National Institutes of Health (\url{www.fnih.org}). The grantee organization is the Northern California Institute for Research and Education, and the study is coordinated by the Alzheimer’s Therapeutic Research Institute at the University of Southern California. ADNI data are disseminated by the Laboratory for Neuro Imaging at the University of Southern California.}

%
\ethics{The AP-HP obtained the authorization of the CNIL (Commission Nationale de l’informatique et des Libertés, the French regulatory body for data
collection and management) in 2017 to share data for research purposes in compliance with the MR004 reference methodology (Daniel and Salamanca,
2020). The MR004 reference controls data processing for the purpose of studying, evaluating and/or researching that does not involve human persons (in the sense of not involving an intervention or a prospective collection of research data in patients that would not be necessary for clinical evaluation, but which allows retrospective use of data previously acquired in patients). The goals of the clinical data warehouse are the development of decision support algorithms, the support of clinical trials and the promotion of multi-centre studies. According to French regulation, and as authorised by the CNIL, patients’ consent to use their data in the projects of the CDW
can be waived as these data were acquired as part of the clinical routine care of the patients. At the same time, AP-HP committed to keep patients
updated about the different research projects of the clinical data warehouse through a portal on the internet6 and individual information is systematically provided to all the patients admitted to the AP-HP. In addition, a retrospective information campaign was conducted by the AP-HP in 2017: it involved around 500,000 patients who were contacted by e-mail and by
postal mail to be informed of the development of the CDW. Accessing the data is possible with the following procedure. A detailed project must be submitted to the Scientific and Ethics Board of the AP-HP. If the project participants are external to AP-HP, they have to sign a contract with the Clinical Research and Innovation Board (Direction de la Recherche Clinique
et de l’Innovation). The project must include the goals of the research, the different steps that will be pursued, a detailed description of the data needed,
of the software tools necessary for the processing, and a clear statement of the public health benefits. Once the project is approved, the research team
is granted access to the Big Data Platform (BDP), which was created by a sub-department of the IT of the AP-HP. The BDP is a platform internal
to the AP-HP where data are collected and that external users can access to perform all their analyses, in accordance with the CNIL regulation. It is
strictly forbidden to export any kind of data and each user can access only a workspace that is specific to their project. Each person of the research team
can access the BDP with an AP-HP account after two-factor authentication. If the research team includes people that are not employed by the AP-HP,
a temporary account associated to the project is activated. The project on which the proposed work is based is called APPRIMAGE, it is led by the
ARAMIS team (current AP-HP PI: Didier Dormont; initial AP-HP PI: Anne Bertrand, deceased March 2nd 2018) at the Paris Brain Institute and it was approved by the Scientific and Ethics Board of the AP-HP in 2018 (Bottani, 2022).}

\coi{Competing financial interests related to the present article: none to disclose for all authors. 

Competing financial interests unrelated to the present article: OC reports having received consulting fees from AskBio (2020), having received fees for writing a lay audience short paper from Expression Santé (2019). 
Members from his laboratory have co-supervised a PhD thesis with myBrainTechnologies (2016-2019) and with Qynapse (2017-present). OC’s spouse is an employee and holds stock-options of myBrainTechnologies (2015-present). 
O.C. holds a patent registered at the International Bureau of the World Intellectual Property Organization (PCT/IB2016/0526993, Schiratti J-B, Allassonniere S, Colliot O, Durrleman S, A method for determining the temporal progression of a biological phenomenon and associated methods and devices) (2017).}

\data{All the data that support the findings of this study, except the ones from the clinical data warehouse, are openly available. 

The Alzheimer's Disease Neuroimaging Initiative (ADNI) data are publicly and freely available from the \href{http://adni.loni.usc.edu/data-samples/access-data/}{http://adni.loni.usc.edu/data-samples/access-data}. 

The MSSEG and MNI BITE data can be found at: \href{https://portal.fli-iam.irisa.fr/msseg-challenge/}{https://portal.fli-iam.irisa.fr/msseg-challenge} and \href{https://nist.mni.mcgill.ca/bite-brain-images-of-tumors-for-evaluation-database/}{https://nist.mni.mcgill.ca/bite-brain-images-of-tumors-for-evaluation-database}.

All the experiments performed in this study were done using the ClinicaDL software \citep{thibeau-sutreClinicaDLOpensourceDeep2021}. The repository is accessible on GitHub: \url{https://github.com/aramis-lab/clinicaDL}.}

\section*{APPRIMAGE Study Group}
\label{sec:study-group}

\noindent Olivier Colliot, Ninon Burgos, Simona Bottani, Sophie Loizillon {$^{1}$} \\
Didier Dormont {$^{1,2}$}, Stéphane Lehéricy{$^{2, 21, 22}$}, Samia Si Smail Belkacem, Sebastian Str\"{o}er {$^{2}$}\\
Nathalie Boddaert {$^{3}$} \\
Farida Benoudiba, Ghaida Nasser, Claire Ancelet, Laurent Spelle {$^{4}$}\\
Aur\'{e}lien Maire, St\'{e}phane Br\'{e}ant, Christel Daniel, Martin Hilka, Yannick Jacob, Julien Dubiel, Cyrina Saussol, Rafael Gozlan {$^{19}$}\\
Florence Tubach, Jacques Ropers, Antoine Roz\`{e}s, Camille Nevoret {$^{20}$}\\
Hubert Ducou-Le-Pointe{$^{5}$}, Catherine Adamsbaum{$^{6}$}, Marianne Alison{$^{7}$}, Emmanuel Houdart{$^{8}$}, Robert Carlier {$^{9,17}$}, Myriam Edjlali{$^{9}$}, Betty Marro{$^{10,11}$}, Lionel Arrive{$^{10}$}, Alain Luciani{$^{12}$}, Antoine Khalil{$^{13}$} , Elisabeth Dion{$^{14}$}, Laurence Rocher{$^{15}$}, Pierre-Yves Brillet{$^{16}$} , Paul Legmann, Jean-Luc Drape {$^{18}$}\\

\begin{small}
\noindent $^{1}$ Sorbonne Universit\'{e}, Institut du Cerveau ‐ Paris Brain Institute, Inserm, CNRS, AP-HP, H\^{o}pital de la Piti\'{e} Salp\^{e}tri\`{e}re, Inria, Aramis project-team, F-75013, Paris, France \\
$^{2}$  AP-HP, H\^{o}pital de la Piti\'{e} Salp\^{e}tri\`{e}re, Department of Neuroradiology, F-75013, Paris, France \\
$^{3}$  AP-HP, H\^{o}pital Necker, Department of Radiology, F-75015, Paris, France \\
$^{4}$  AP-HP, H\^{o}pital Bic\^{e}tre, Department of Radiology, F-94270, Le Kremlin-Bic\^{e}tre, France \\
$^{5}$  AP-HP, H\^{o}pital Armand-Trousseau, Department of Radiology, F-75012, Paris, France \\
$^{6}$  AP-HP, H\^{o}pital Bic\^{e}tre, Department of Pediatric Radiology, F-94270, Le Kremlin-Bic\^{e}tre, France \\
$^{7}$  AP-HP, H\^{o}pital Robert-Debr\'{e}, Department of Radiology, F-75019, Paris, France \\ 
$^{8}$  AP-HP, H\^{o}pital Lariboisi\`{e}re , Department of Neuroradiology, F-75010, Paris, France \\
$^{9}$  AP-HP, H\^{o}pital Raymond-Poincar\'{e}, Department of Radiology, F-92380, Garches, France \\
$^{10}$ AP-HP, H\^{o}pital Saint-Antoine, Department of Radiology, F-75012, Paris, France \\
$^{11}$ AP-HP, H\^{o}pital Tenon, Department of Radiology, F-75020, Paris, France \\
$^{12}$ AP-HP, H\^{o}pital Henri-Mondor, Department of Radiology, F-94000, Cr\'{e}teil, France \\
$^{13}$ AP-HP, H\^{o}pital Bichat, Department of Radiology, F-75018, Paris, France \\
$^{14}$ AP-HP, H\^{o}pital H\^{o}tel-Dieu, Department of Radiology, F-75004, Paris, France \\
$^{15}$ AP-HP, H\^{o}pital Antoine-B\'{e}cl\`{e}re, Department of Radiology, F-92140, Clamart, France \\
$^{16}$ AP-HP, H\^{o}pital Avicenne, Department of Radiology, F-93000, Bobigny, France \\
$^{17}$ AP-HP, H\^{o}pital Ambroise Par\'{e}, Department of Radiology, F-92100 104, Boulogne-Billancourt, France \\ 
$^{18}$ AP-HP, H\^{o}pital Cochin, Department of Radiology, F-75014, Paris, France \\ 
$^{19}$ AP-HP, WIND department, F-75012, Paris, France \\
$^{20}$ AP-HP, Unit\'{e} de Recherche Clinique, H\^{o}pital de la Piti\'{e} Salp\^{e}tri\`{e}re, Department of Neuroradiology, F-75013, Paris, France  \\
$^{21}$ ICM, Centre de NeuroImagerie de Recherche – CENIR, Paris, France  \\
$^{22}$ Sorbonne Université, Institut du Cerveau – Paris Brain Institute, Inserm, CNRS, AP-HP, Hôpital de la Pitié Salpêtrière, F-75013, Paris, France \\
\end{small}

\bibliography{ref}


\clearpage

\renewcommand{\theHsection}{A\Alph{section}}

\renewcommand\theHfigure{A{\arabic{figure}}} 
\renewcommand\thefigure{A{\arabic{figure}}} 
\setcounter{figure}{0} 
\renewcommand\theHtable{A\arabic{table}} 
\renewcommand\thetable{A\arabic{table}} 
\setcounter{table}{0}   

\appendix

\section{Supplementary figures and tables}
\label{sec:suppl_fig_tab}

\subsection*{Manual annotation}
\begin{table}[ht]
   \centering
   \caption{Weighted Cohen's kappa between the two manual annotators for the three types of artefacts (motion, poor contrast, noise). More in~\citep{bottaniAutomaticQualityControl2021}.}
   \begin{tabular}{cccc}
   \hline
        & Motion (0 vs 1 vs 2)   & Noise (0 vs 1 vs 2)   & Contrast (0 vs 1 vs 2)  \\ \hline
       Kappa Score & 0.68 & 0.70 & 0.79 \\ \hline
   \end{tabular}
   \label{suptab:kappa}
\end{table}

\FloatBarrier

\subsection*{Direct and indirect tier classification}

\begin{figure}[ht!]
    \centering
    \caption{Comparison of the two workflows of the transfer learning framework. In the direct tier classification, corrupted research MRIs with synthetic artefacts (motion, noise and poor contrast) are used to pre-train two CNNs to directly detect the tier of the images (moderate quality MRIs detection and bad quality MRIs detection). Models are then fine-tuned on routine clinical data relying on the manual annotation of these artefacts for 5000 MRIs. In the indirect tier classification, three models are pre-trained for each type of artefacts. Then the three models are fine-tuning on clinical data with real artefacts. Finally, the quality tiers are then determined using the outputs of the artefact-specific models.}
    \includegraphics[width=1\linewidth]{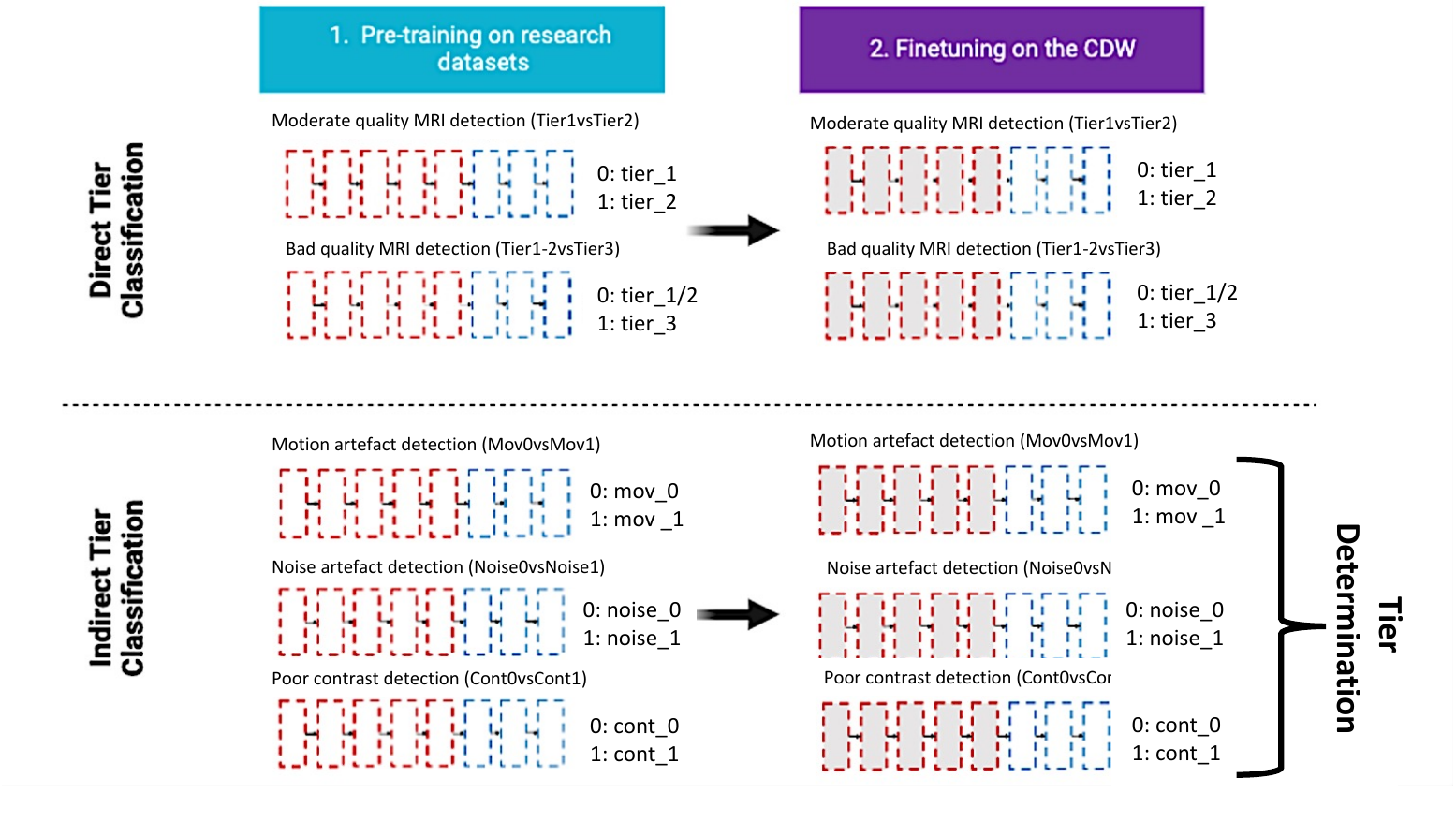}
    \label{subfig:direct_indirect_workflow}
\end{figure}

\subsection*{Motion quantification}
\begin{figure}[ht!]
   \centering
   \caption{Boxplot of the average edge strength (AES) and Tenengrad measure for the clinical routine dataset ordered by motion severity.}
   \includegraphics[width=\linewidth]{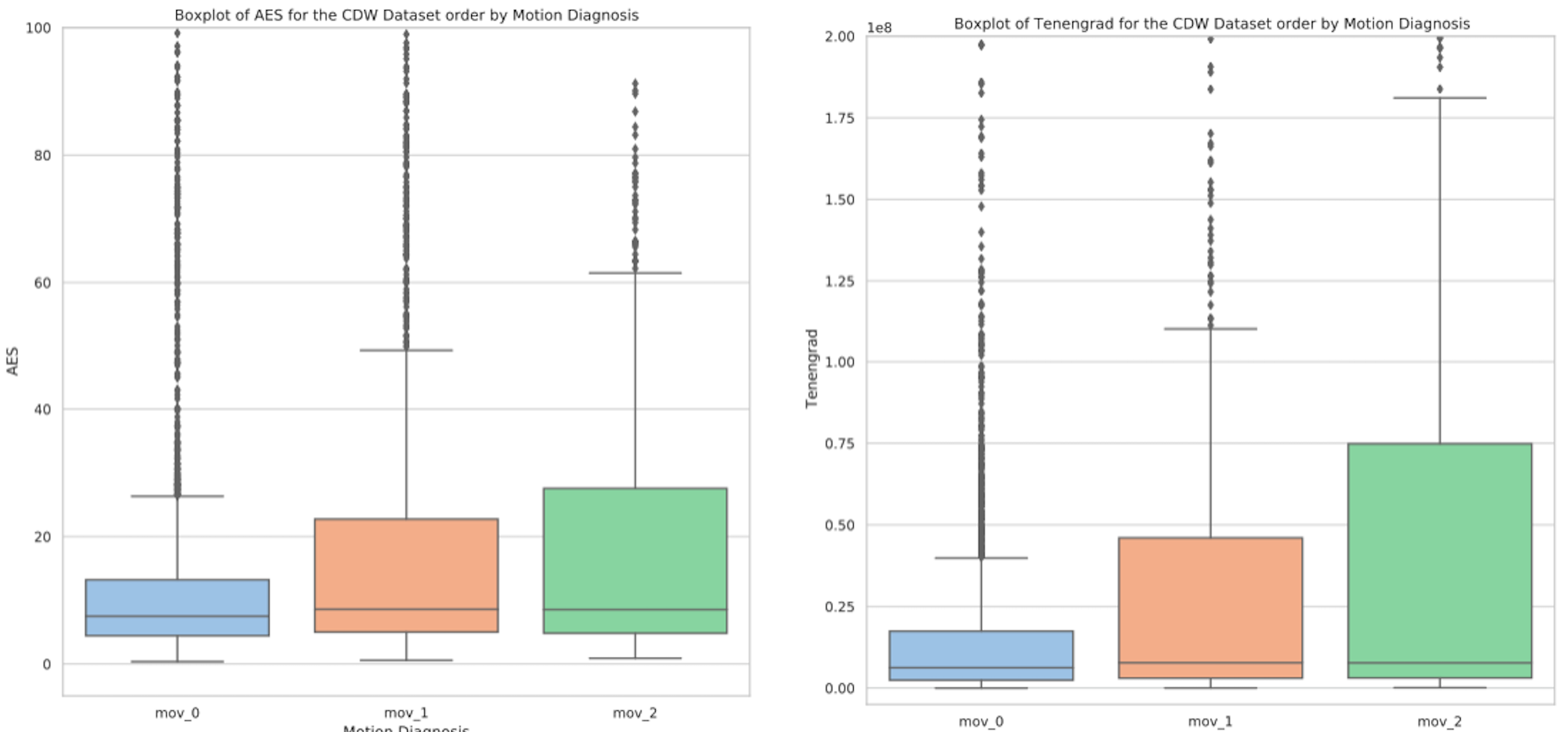}
   \label{supfig:motion_compute}
\end{figure}

In their paper~\cite{eichhorn2022evaluating} show that SSIM and PSNR are the metrics that correlate the best with radiological assessment. To compute these metrics, we need a reference MRI, which is not the case in our clinical dataset. Their work suggests that the average edge strength and the Tenengrad measure are reference-free metrics the most strongly correlated with motion. Therefore, we computed these metrics on our routine clinical dataset for which we had manual annotations (c.f. Figure~\ref{supfig:motion_compute}).

\begin{figure}[ht!]
   \centering
   \caption{Boxplot of the average edge strength (AES) ordered by motion and by contrast diagnosis.}
   \includegraphics[width=1\linewidth]{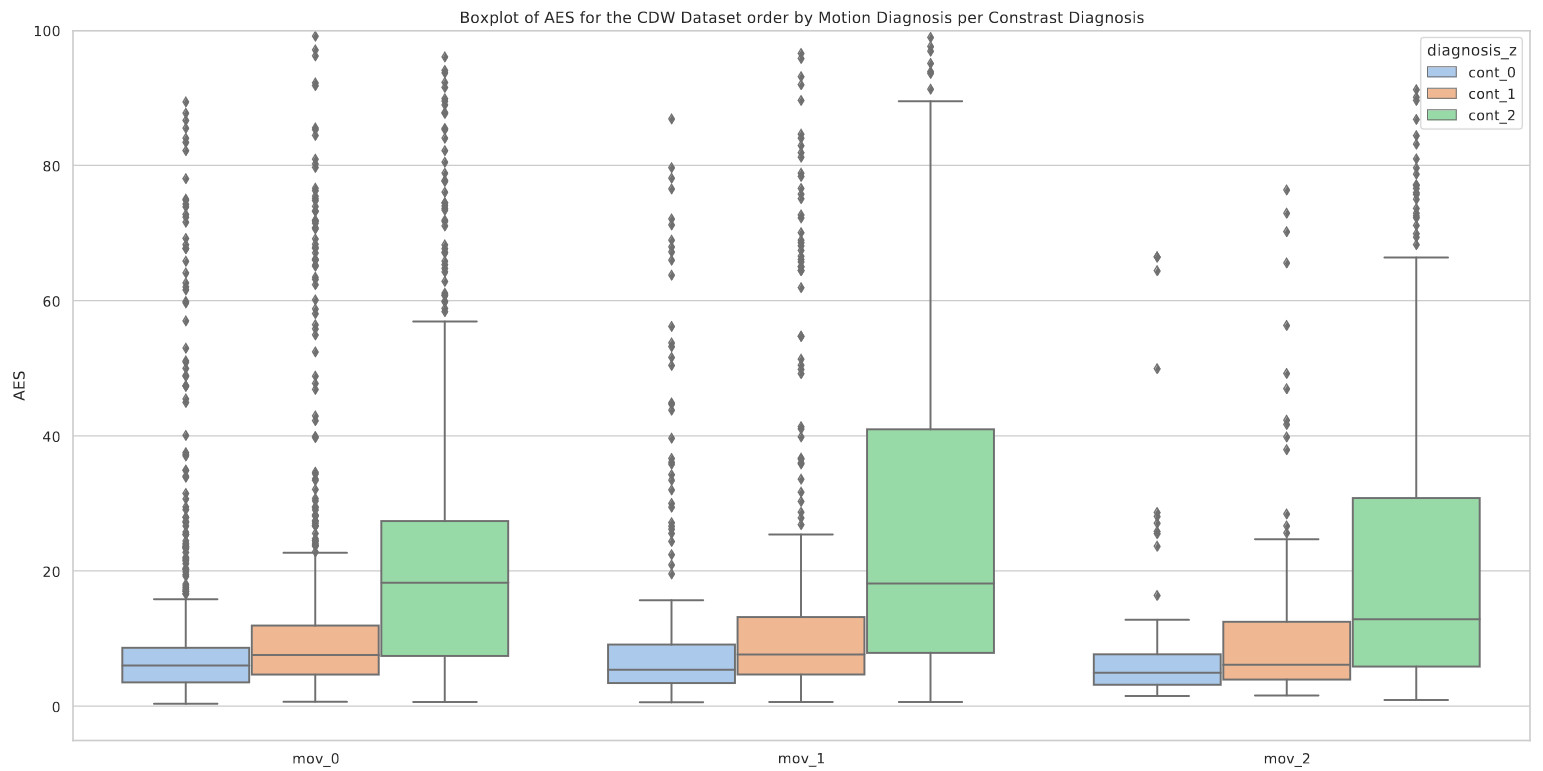}
   \label{supfig:aes_contrast}
\end{figure}

\begin{figure}[ht!]
   \centering
   \caption{Boxplot of Tenengrad ordered by motion and by contrast diagnosis.}
   \includegraphics[width=1\linewidth]{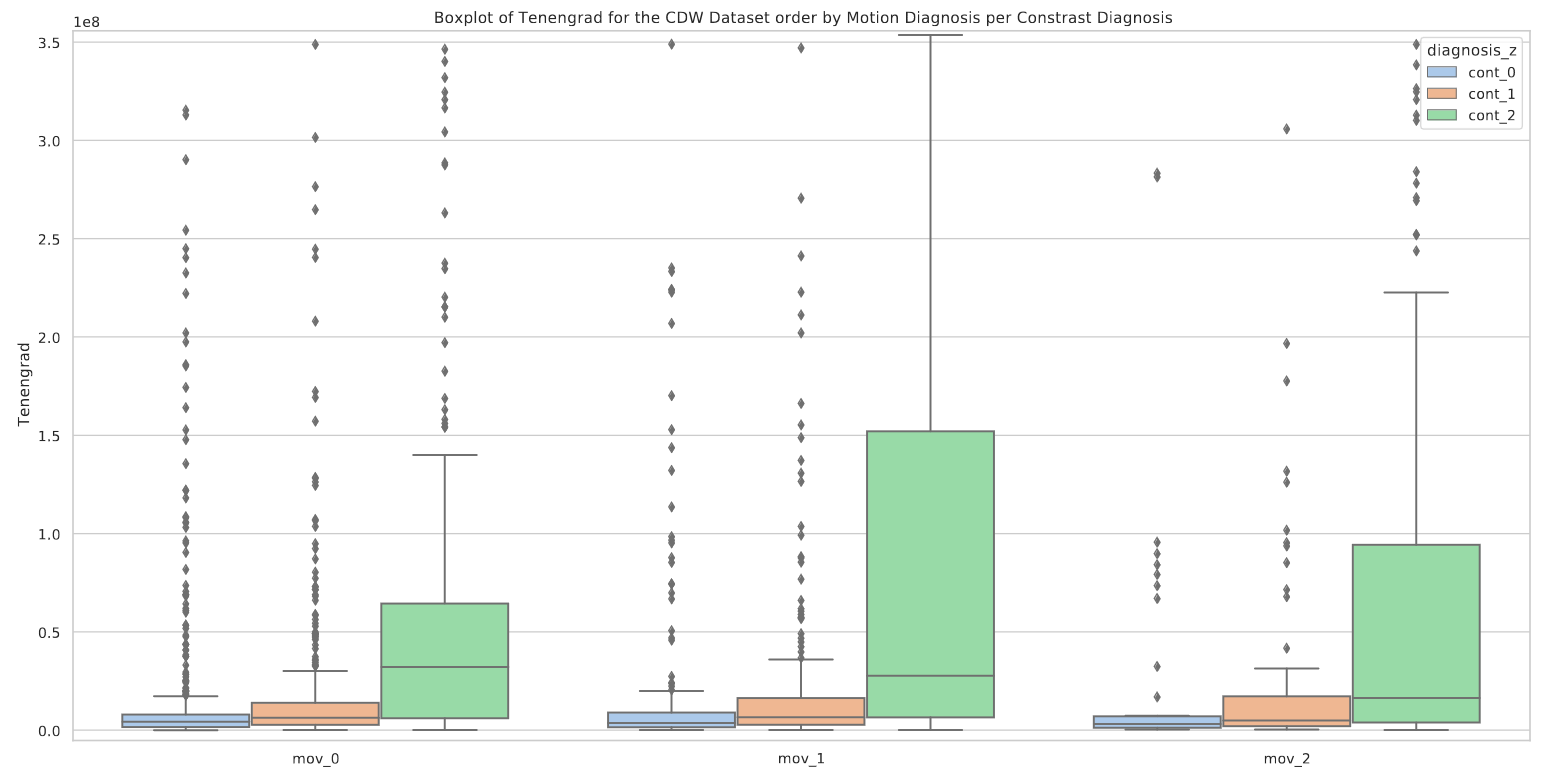}
   \label{supfig:tgn_contrast}
\end{figure}

Neither the AES nor the Tenengrad measure showed a significant correlation with the manual annotations. Motion quantification is a complex problem, mainly due to its sensitivity to many cofactors such as contrast. Each of these metrics is sensitive to the quality of the image contrast as underlined in Figures~\ref{supfig:aes_contrast} and \ref{supfig:tgn_contrast}. A robust quantification of the motion is therefore still an open problem.

\FloatBarrier
\clearpage

\subsection*{Data splits}

\begin{table}[ht] 
\centering
\caption{Distribution of the training, validation and test sets separately for the moderate artefact detection task using the research dataset comprising images from the ADNI, MSSEG and MNI BITE databases.\\ }
\begin{tabular}{|c|c||cc|cc|cc|}
\hline
 & Task & \multicolumn{2}{c|}{Motion detection} & \multicolumn{2}{c|}{Contrast detection}  & \multicolumn{2}{c|}{Noise detection}  \\ \hline
                   & Label         & Mov0           & Mov1     &  Cont0           & Cont1       & Noise0  & Noise1 \\ \hline
\multirow{3}{*}{\begin{turn}{90} \small Train \end{turn}}& ADNI           & 2859                & 953             & 2859                & 953               & 2859                & 953                           \\ 
& MSSEG           & 117                & 39              & 117                & 39              & 117                & 39                           \\ 
& MNI           & 114                & 38              & 114                & 38              & 114                & 38             \\ \hline

\multirow{3}{*}{\begin{turn}{90} \small Val \end{turn}}& ADNI           & 294                & 98             & 294                            & 98  & 294                            & 98        \\ 
& MSSEG           & 30                             & 10 &30                          & 10 & 30                            & 10           \\ 
& MNI           & 15                            & 5 & 15                           & 5  & 15                & 5                          \\ \hline

\multirow{3}{*}{\begin{turn}{90} \small Test \end{turn}}& ADNI           & 162                & 54             & 162                & 54             & 162                & 54         \\ 
& MSSEG           & 15                & 5            & 15               & 5           & 15                & 5                  \\ 
& MNI           & 6                & 2          & 6                & 2          & 6                & 2                       \\ \hline

\end{tabular}
\label{subtab:distribution_research}
\end{table}

\begin{table}[ht]
\centering
\caption{Distribution of the training, validation and test sets separately for the severe artefact detection task using the research dataset comprising images from the ADNI, MSSEG and MNI BITE databases.\\ }
\begin{tabular}{|c|c||cc|cc|cc|}
\hline
 & Task & \multicolumn{2}{c|}{Motion detection} & \multicolumn{2}{c|}{Contrast detection}  & \multicolumn{2}{c|}{Noise detection}  \\ \hline
                   & Label         & Mov0/1          & Mov2     &  Cont0/1          & Cont2   & Noise0/1          & Noise2  \\ \hline
\multirow{3}{*}{\begin{turn}{90} \small Train \end{turn}}& ADNI           & 2859/953             & 953 & 2859/953             & 953  & 2859/953             & 953               \\ 
& MSSEG           & 117/39             & 39 & 117/39             & 39  & 117/39             & 39               \\ 
& MNI           & 114/38             & 38 & 114/38             & 38  & 114/38             & 38               \\ \hline

\multirow{3}{*}{\begin{turn}{90} \small Val \end{turn}}& ADNI           & 294/98             & 98 & 294/98             & 98  & 294/98             & 98        \\ 
& MSSEG           & 30/10             & 10 &30/10             & 10 & 30/10             & 10           \\ 
& MNI           & 15/5             & 5 & 15/5             & 5  & 15/5             & 5               \\ \hline

\multirow{3}{*}{\begin{turn}{90} \small Test \end{turn}}& ADNI           & 162/ 54             & 54 & 162/54             & 54  & 162 /54             & 54        \\ 
& MSSEG           & 15 / 5             & 5 & 15/ 5             & 5 & 15/5             & 5           \\ 
& MNI           & 6/ 2             & 2 & 6/2             & 2  & 6 / 2             & 2               \\ \hline

\end{tabular}
\label{subtab:distribution_research_severe}
\end{table}

\begin{table}[ht] 
\centering
\caption{Distribution of the training, validation and test sets separately for the severe artefact detection task using the clinical dataset comprising MRIs with 'real' artefacts.\\ }
\begin{tabular}{|c||cc|cc|cc|}
\hline
  & \multicolumn{2}{c|}{Motion detection} & \multicolumn{2}{c|}{Contrast detection}  & \multicolumn{2}{c|}{Noise detection}  \\ \hline
                            & Mov0/1          & Mov2     &  Cont0/1          & Cont2   & Noise0 & Noise1/2  \\ \hline
  Train      &  2540& 379& 1876& 1055&1949& 982\\ 
         Validation & 647& 94&  458&  271& 496& 233\\ 
          Test &  328& 57& 238 &147 & 258& 127 \\ \hline 

\end{tabular}
\label{subtab:distribution_clinical_severe}
\end{table}

 \begin{table}[ht] 
\centering
\caption{Distribution of the training, validation and test sets separately for the moderate artefact detection task using the clinical dataset comprising MRIs with 'real' artefacts.\\ }
\begin{tabular}{|c||cc|cc|cc|}
\hline
 & \multicolumn{2}{c|}{Motion detection} & \multicolumn{2}{c|}{Contrast detection}  & \multicolumn{2}{c|}{Noise detection}  \\ \hline
                           & Mov0          & Mov1     &  Cont0          & Cont1   & Noise0 & Noise1 \\ \hline
        Train & 1681 & 859 & 1111 & 765 & 1949 & 865 \\ 
        Validation& 428 & 219 & 273 &  185&  496 & 232  \\ 
        Test &210 & 118  &135 & 103  &258& 125\\ \hline

\end{tabular}
\label{subtab:distribution_clinical_moderate}
\end{table}

\FloatBarrier

\subsection*{Failure cases}

\begin{figure}[ht!]
   \centering
   \caption{Examples of failures in the detection of noise, motion and poor contrast artefacts. For each artefact, we display an example of a false negative and a false positive.}
   \includegraphics[width=\linewidth]{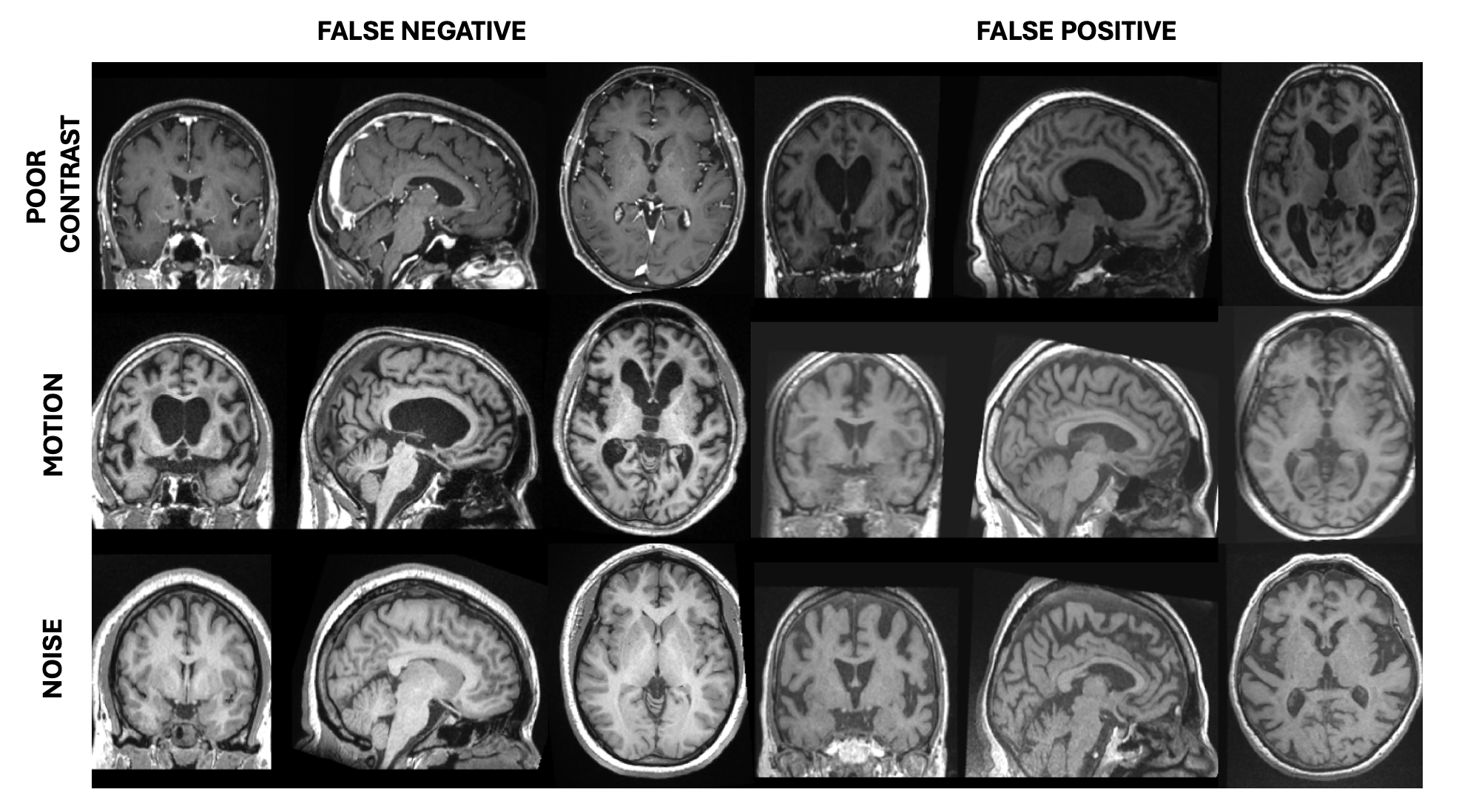}
   \label{supfig:failure_cases}
\end{figure}

\FloatBarrier

\end{document}